\documentclass[10pt,letterpaper,twocolumn]{article} 
\usepackage{ol2}
\usepackage[draft]{hyperref}
\usepackage{amsmath}

\begin{document}
\twocolumn[ 

OPTICS LETTERS, {\bf 34}, 12, pp. 1780-1782 (2009), DOI:10.1364/OL.34.001780

\title{Radially Polarized Conical Beam from an Embedded Etched Fiber}

\author{Djamel Kalaidji $^{1,2}$, Michel Spajer $^{1,*}$, Nad\`ege Marthouret $^1$ \\ and Thierry Grosjean $^1$}

\address{$^1$ Institut FEMTO-ST, Universit\' e de Franche-Comt\' e, UMR 6174 CNRS\\
D\' epartement d'Optique, 16 route de Gray, 25030 Besan\c con cedex, France \\
$^2$ Universit\' e Abou Bekr Belkaid, Facult\' e des Sciences, \\ D\' epartement de Physique, BP 119, 13000 Tlemcen, Alg\' erie \\
$^*$ Corresponding author: michel.spajer@univ-fcomte.fr
}

\begin{abstract}
We propose a method for producing a conical beam based on the lateral refraction of the $TM_{01}$ mode from a two-mode fiber after chemical etching of the cladding, and for controlling its radial polarization. The whole power of the guided mode is transferred to the refracted beam with low diffraction. Polarization control by a series of azimuthal detectors and a stress controller affords the transmission of a stabilized radial polarization through an optical fiber. A solid component usable for many applications has been obtained.
\end{abstract}


] 

\maketitle

\noindent Radially polarized and conical beams have proved their interest, among many other applications \cite{zhan09}, both in confocal \cite{sheppard04} or near-field microscopy \cite{grosjean03}. Different means of shaping free propagating beams, guided or laser mode by using birefringent components \cite{erdelyi08} or axicons \cite{li06} have been proposed. As an alternative technique, the use of the $TM_{01}$ propagation mode of optical fibers \cite{grosjean03} does not depend of a specific laser source and allows remote sensing or remote machining. A first difficulty comes from the existence of the fundamental $HE_{11}$ mode whatever be the index profile of the core \cite{ramachandran05}. Several methods has been proposed for selective launching of the linearly polarized mode $LP_{11}$ since the 80's, based on phase masks \cite{facq84,grosjean05}. To avoid any residual coupling on the $HE_{11}$ mode, the most reliable method of mode filtering at the fiber output is lateral refraction \cite{shaklan91}. The second difficulty consists in obtaining a pure $TM_{01}$ mode from the initial $LP_{11}$ combination and in the coupling between $HE_{21}$, $TM_{01}$ and $TE_{01}$. It has been proved that the transformation of $LP_{11}$ into $TM_{01}$ mode can be achieved by a proper stress generation on the fiber \cite{grosjean05} with a longer fiber and a shorter stressed segment than mentioned in ref.\cite{mcgloin98}. Other authors address this mode conversion in terms of vortex inversion \cite{alexeyev08}. Recent papers open the way to radial polarization maintaining fibers by using a specific index profile which lifts polarization degeneracy \cite{ramachandran05}. The object of this paper is to assess the quality of the circular refracted beam that can be expected after chemical etching of a standard step index fiber, the efficiency of the polarization control by a series of azimuthal detectors and of the mode conversion with an arbitrary long and curved fiber.

\begin{figure}[!htbp]
        \includegraphics[width=7.0cm]{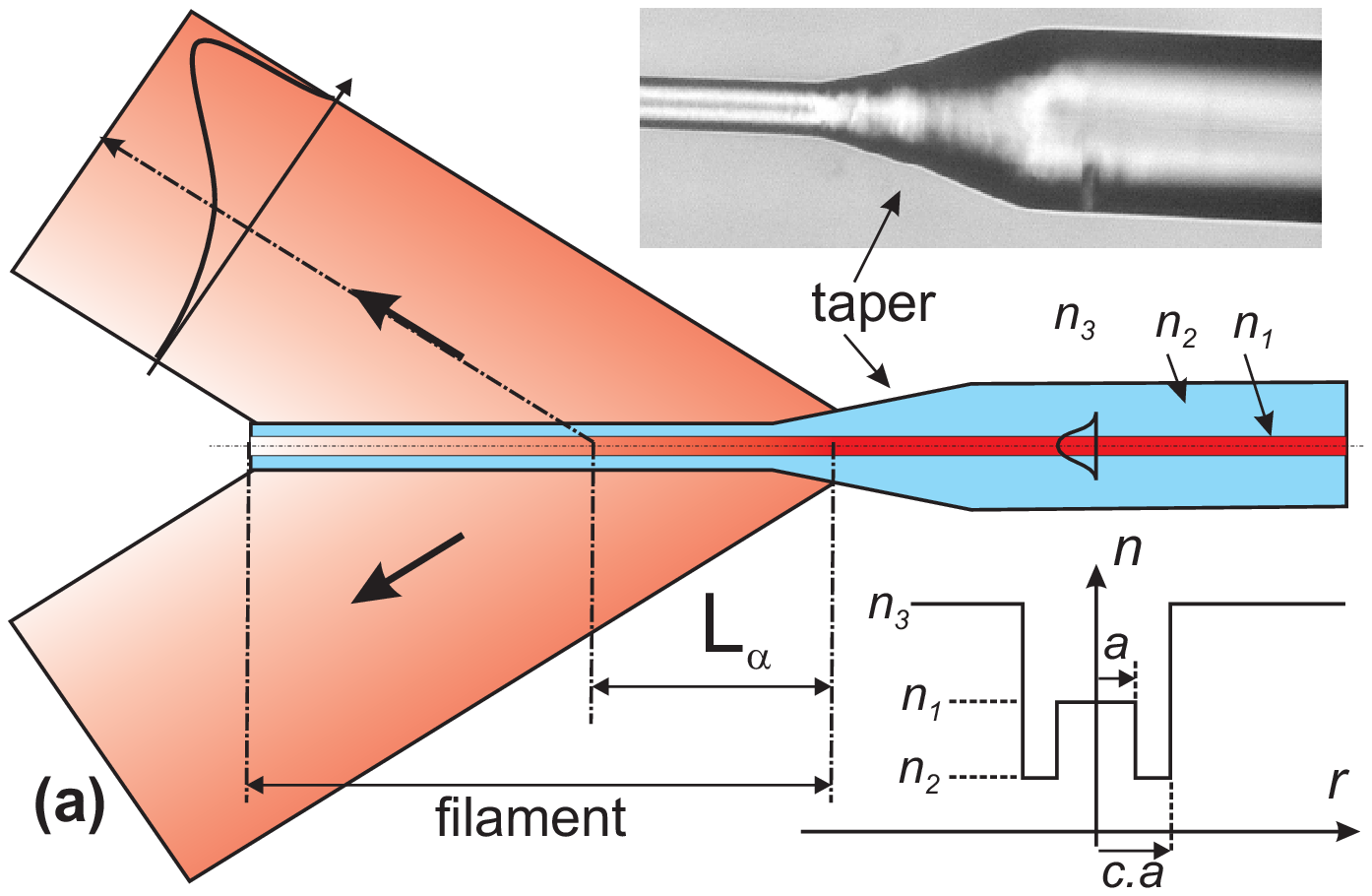}
        \includegraphics[width=7.0cm]{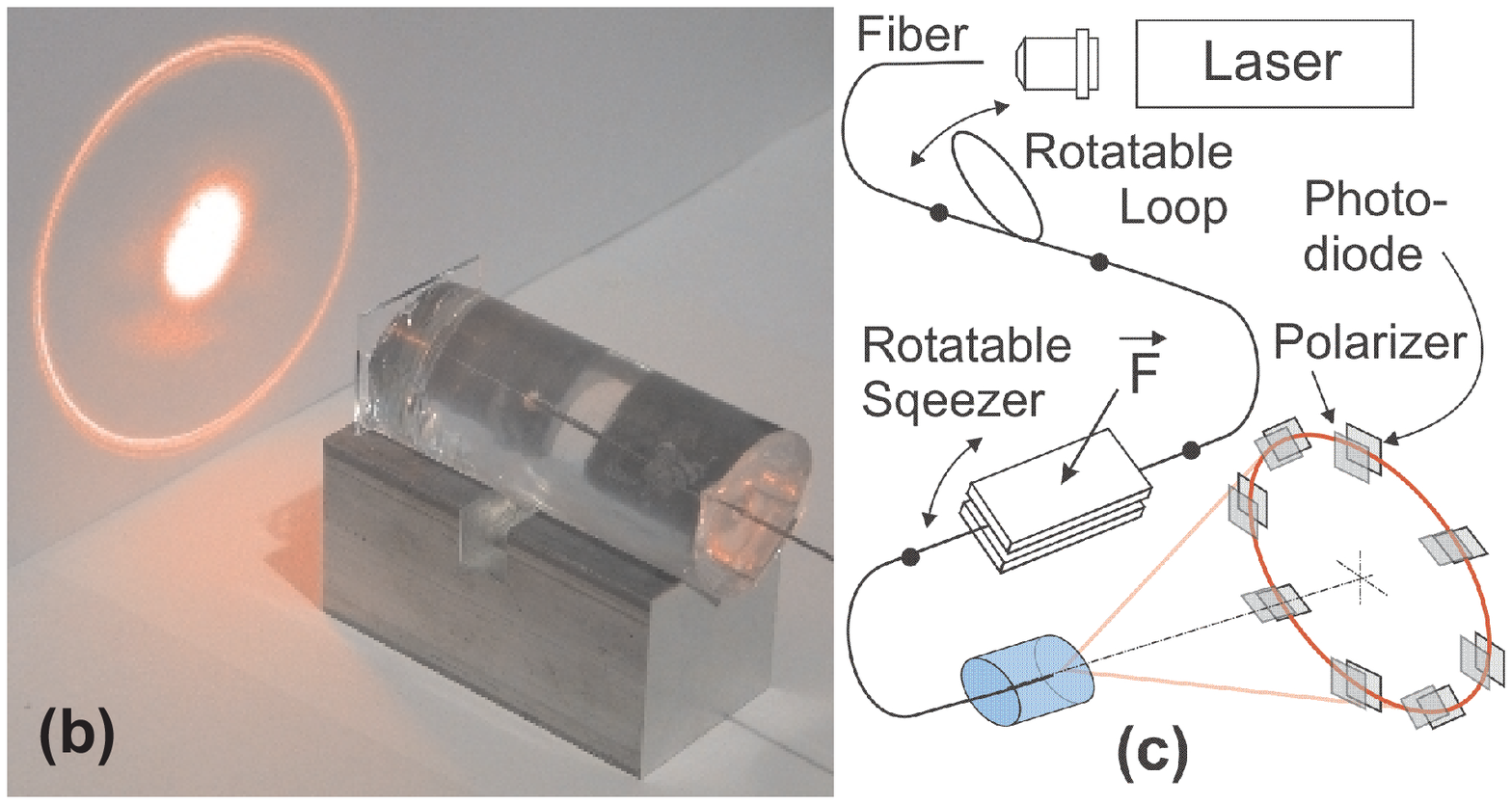}
        \caption{(a)Principle of the refraction of a guided mode from an etched fiber into a high index medium. (b) Refraction in a solid polyester resin. (c) Experimental set-up with 8 detectors: a small polarizer orthoradially oriented is placed in front of each detector.}
       \label{Refraction}
\end{figure}

The lateral refraction of any guided mode of an optical fiber can be obtained by transforming the weakly guiding fiber into a leaky waveguide. This is easily done by reducing the cladding diameter in a buffered HF solution and immersing the etched fiber in a high index liquid. The typical etching speed is about 1 micron/minute in $40 \%$ HF and can be reduced in diluted acid to improve the precision on the final diameter. The etched segment (filament) of the fiber can be modelled by a 3-medium (double-cladding) waveguide connected to the original fiber by  a tapered transition (Fig.\ref{Refraction}a).
Besides, the refracted beam must enter a high numerical aperture objective to be focused elsewhere. To this aim the etched filament must be as short as possible.
The vaseline oil we use as the immersion liquid were successfully replaced by a polyester resin ({\it Soloplast GTS PRO} from {\it Vosschemie}) used for inclusions (Fig.\ref{Refraction}b).
Its obvious advantage is the fabrication of a solid component where the filament can be close to the end face. This facilitates the use of an objective with a short working distance, but the adhesion of the polymer to the filament and its long term stability must be improved. Therefore most of our experiments were done with oil. In the example of Fig.\ref{Refraction}b the filament is too short to allow a complete refraction of $LP_{01}$ and $LP_{11}$ modes and to avoid a direct emission of the residual power from its extremity.

The leakage rate of this device depends on the residual thickness of the cladding along the filament. The beam profile depends also on the filament length and the taper shape. Considering the index profile of Fig.\ref{Refraction}a, we calculated the parameters of the leaky mode as a function of the cladding thickness. The dispersion curve $u(V)$ has already been studied in reference \cite{kawakami75} in the case of guided modes ($n_3<n_1$). We use the same equation in the case $n_3>n_1$ where all modes are refracted:
\begin{eqnarray}
 \frac{J_{n}(u)}{u J_{n+1}(u)}-\frac{K_{n}(w)}{w K_{n+1}(w)}=-(\frac{J_{n}(u)}{u J_{n+1}(u)}+\frac{I_{n}(w)}{w I_{n+1}(w)}) . \nonumber \\
 \frac{K_{n+1}(cw)I_{n+1}(w)}{K_{n+1}(w)I_{n+1}(cw)} .
 \frac{\frac{K_{n}(cw)}{cw K_{n+1}(cw)}-\frac{H_{n}(cW)}{cW H_{n+1}(cW)}}{\frac{I_{n}(cw)}{cw I_{n+1}(cw)}-\frac{H_{n}(cW)}{cW H_{n+1}(cW)}}
\label{eq dispersion}
\end{eqnarray}
where :
\begin{eqnarray}
u^{2}+w^{2}=V^{2} & & u^{2}+W^{2}=V^{2} \frac{n_1-n_3}{n_1-n_2}
\label{eq uwV}
\end{eqnarray}

\begin{figure}[!htbp]
        \includegraphics[width=7.5cm]{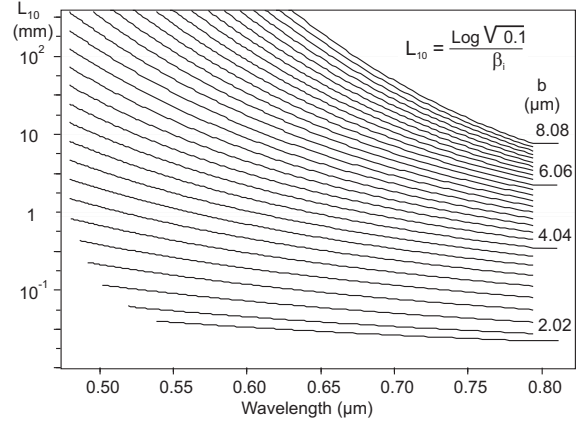}
        \caption{Efficient length of the filament vs. wavelength and residual cladding thickness. Fiber parameters measured by XLIM laboratory: a=2.02 $\mu m$, $n_1$=1.47, $n_2$=1.4622, $\Delta=(n_1 - n_2)/n_1$=0.0053. Oil: $n_3$=1.486}
        \label{Dispersion}
\end{figure}

In these equations, the normalized transverse propagation constants in the core ($u$), cladding ($w$) and
immersion liquid ($W$) are complex, and $W^{2}$ has a negative real part. $V=kn_1a(2 \Delta) ^{1/2}$ is the normalized frequency of the initial single cladding fiber. The attenuation coefficient $\beta _i$, or imaginary part of the propagation constant $\beta$, is derived from the dispersion curves $u(V)$ according to the equation:
\begin{eqnarray}
k = 2 \pi / \lambda  & &  \beta = \sqrt{(n_1 k)^2-(u/a)^2}
\label{eq beta}
\end{eqnarray}
The relevant parameter we preferred to draw (Fig.\ref{Dispersion}) is the efficient length of the filament for different cladding thickness. We define it as the propagation length where the mode power is divided by $10$. Under this condition, $90 \%$ of its power is transferred to the refracted beam. The curves of Fig.\ref{Dispersion} show the important variation of the attenuation in the visible range and the extreme precision in the etching time that is necessary to ensure a precise value of the attenuation for a given wavelength. Therefore, the section of the etched cladding must be very regular: perfectly circular and concentric to the core.

\begin{figure}[!htbp]
        \includegraphics[width=7.0cm]{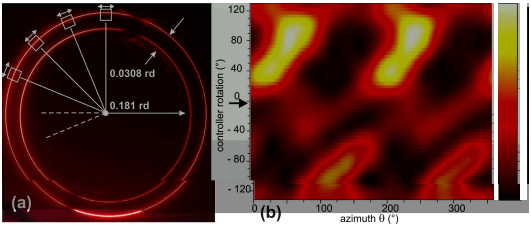}
        \caption{(a) Refracted beams obtained with a 15 mm filament. Detector positions and analyzer orientation are marked. (b) Gray level encoding of the azimuthal intensity distribution vs. the rotation of the sqeezer.}
        \label{Halos}
\end{figure}

After some meters of propagation the $LP_{11}$ mode which is initially launched into the fiber is replaced by an arbitrary combination of $HE_{21}$, $TM_{01}$ and $TE_{01}$ modes. Polarization modification results from mode beating and from mode coupling induced by local perturbation. Polarization stabilization can be achieved by the proper compensation between these two phenomena or by the fabrication of a fiber that maintains this polarization. A complete description of this mode combination supposes a measurement of the state of polarization (SOP) at a minimum of 8 azimuthal positions (Fig.\ref{Refraction}c). To have a better sampling of the azimuthal intensity variations, we chose a 16-detector array. The SOP can be measured with a rotating polarizer, but our main interest is the weight of the $TM_{01}$ mode. Therefore we limit our measurements to the orthoradial component of the electric field with a series of fixed polarizers (Fig.\ref{Halos}a).

The use of Lefevre loops \cite{lefevre80} to control the polarization of the fundamental mode and their equivalence to fractional wave plates cannot be simply applied to $LP_{11}$ mode family: such a birefringent fiber segment is more likely to be a coupling device between $TM_{01}$ and $HE_{21}$ or $TE_{01}$ and $HE_{21}$, as far as the birefringence $\delta \beta$ is comparable to the difference $\Delta \beta$ between modes \cite{snyder83}.
A rotatable loop of adjustable diameter (20-60 mm) has been introduced on the fiber without obvious influence on the result. A commercial device of shorter dimension has experimentally proved to be a better polarization controller \cite{grosjean05}: we used an in-line Babinet-Soleil compensator made of a rotatable fiber squeezer, 23 mm long ({\it Newport}, ref. {\it F-POL-IL}). In our experiment, its adjustment must ensure a minimum signal for all of the 16 detectors.
Further studies will show if the theoretical analysis of Alexeyev et al. \cite{alexeyev08} can explain the complete or limited polarization control with an arbitrarily curved fiber.

Fig.\ref{Halos}a shows the refracted beams of both fundamental and second order modes from the oil tank where the fiber is immersed. The filament length is about $15 mm$: both propagating mode (fundamental $LP_{01}$ and $LP_{11}$) are completely refracted and give rise to a distinct ring. No direct emission is visible in the center of the field. The filament is maintained in tension by a spring made of a copper wire, the shadow of which interrupts the two rings. The refracted cones have inhomogeneous azimuthal distributions of intensities (ratio = 7.6 between the left and the right side). Possible explanations are a slight eccentricity of the core in the fiber, a slight default of circularity of the fiber, an inhomogeneity of the etching rate depending on the fiber position in the HF tank, or an inhomogeneity of HF diffusion into the fiber. It is worthwhile to notice that the ring width is not minimum at the fiber extremity but at some distance (typically 60 cm) from the extremity. Such an annular beam can be focused to obtain a Bessel beam for near-field microscopy, mainly with evanescent wave illumination \cite{grosjean03}.

The azimuthal inhomogeneity led us to normalize each of the 16 signals obtained with an orthoradial polarizer by a reference signal which is the maximum value obtained without polarizer by manual adjustment of the controller. An example of this normalization, for a fixed value of the lateral stress, is encoded in gray levels on Fig.\ref{Halos}b: the horizontal axis represents the azimuthal coordinate of the detector and the continuous intensity distribution is calculated from the 16 signals by spline interpolation. The vertical axis represents the rotation angle of the squeezer. The arrow on the left side indicates the position that ensures the minimum signal.

\begin{figure}[!htbp]
        \includegraphics[width=6.0cm]{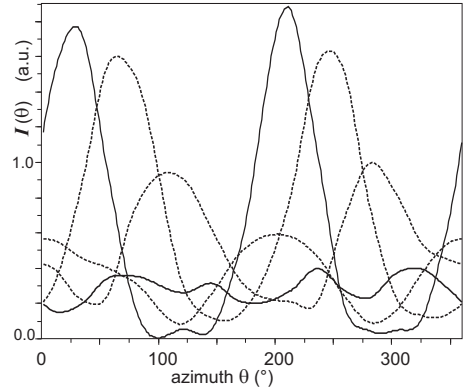}
        \caption{Normalized azimuthal variation of the signal with different adjustments of the stress controller}
        \label{Detection}
\end{figure}

The curves of Fig.\ref{Detection} represent more clearly the azimuthal dependence of intensity for different adjustment of the controller: the plain curves correspond to the maximum and minimum intensity, the dotted one to intermediate adjustments. They are in qualitative accordance with the general form of the signal obtained with radial (or orthoradial) polarizers as a function of azimuth $\theta$:
\begin{equation}
I(\theta) = 1 + \gamma [cos(\theta) cos (\theta _0 - \theta)]^2
\label{eq Theta}
\end{equation}
where $\gamma$ is the visibility of the azimuthal modulation, $\theta _0$ its orientation.
The extinction ratio is not smaller than 0.20 but could be reduced by an optimized mechanical stress applied to the fiber. This extinction were obtained by combining at arbitrary positions along the fiber the commercial squeezer and a rotatable loop of 30 mm diameter (Fig.\ref{Refraction}c).

As a conclusion, lateral refraction of $TM_{01}$ mode is a reliable method to transfer the whole power of the guided mode to a conical beam with perfectly circular cross-section and radial polarization. The transfer efficiency is higher and the diffraction much lower than in a conical beam generated by an axicon. The inhomogeneity of the azimuthal power distribution we observed in the first experiments could be improved by a better control of the chemical process or by alternative methods such as fiber pulling. A series of polarized detectors gives a good control of mode purity. This diagnosis can be extended to other fiber shaping methods and to other kind of waveguide such as photonic fibers. A systematic study of different mechanical stress and/or fiber structures that can enhance the selection of the radial polarization is necessary. Embedment of the fiber extremity in a polymer matrix makes the component easy to handle for a great variety of applications.

The authors thank Jean-Louis Auguste from XLIM laboratory (UMR 6172) of Limoges (France) for his measurements of the fiber index profile.

\end{document}